\documentclass[11pt]{article}
\usepackage{moriond,epsfig}
\usepackage{amsmath}

\def\be{\begin{equation}}
\def\ee{\end{equation}}
\def\bea{\begin{eqnarray}}
\def\eea{\end{eqnarray}}

\begin{document}
\vspace*{4cm}
\title{THE PRECISION REACH OF ATLAS FOR ELECTROWEAK PHYSICS IN THE LOW LUMINOSITY ERA}

\author{ P. J. BELL }

\address{CERN, CH-1211, Geneva 23, Switzerland }

\maketitle\abstracts{
During the first three years of low luminosity operation, the LHC will facilitate a number of precision tests of the Electroweak sector of the Standard Model.
The prospects for measuring the W boson and top quark masses and for probing the charged triple gauge boson couplings at ATLAS are presented here.  In each case the most likely analysis methods are described and the statistical and systematic uncertainties which can be expected are reported.
}

\section{Introduction}

From 2007 the Large Hadron Collider (LHC) at CERN will collide two beams of protons with a centre-of-mass energy of 14~TeV.  The general purpose ATLAS (A Toriodal LHC Apparatus) experiment has been designed to realise the main physics goals of the LHC, namely searches for the Higgs boson and for supersymmetric particles.
 In addition to its discovery potential, however, the LHC will constitute a powerful tool with which to make precision tests within the Standard Model (SM).  The design luminosity of the LHC is $10^{34}$~cm$^{-2}$s$^{-1}$ and even during the initial three year phase of low luminosity running ($2\times10^{33}$~cm$^{-2}$s$^{-1}$) the nominal integrated luminosity per year will be 10~fb$^{-1}$. The reduced effects of pile-up during this period  makes it favourable for performing many precision measurements.  The rates of selected physics processes and the number of events expected per low luminosity year are shown in table~\ref{tab:processes}.  
\begin{table}[t]
\caption{Examples of production rates at the LHC and the corresponding number of events expected per low luminosity year of running $(10~$fb$^{-1})$. For comparison the total statistics available by the time of LHC start-up from other experiments are shown.}
\label{tab:processes}
\vspace{0.4cm}
\begin{center}
\begin{tabular}{|c|c|c|c|}
\hline
Process & LHC Rate~[Hz] & LHC events/year & Total events by 2007\\
\hline
Z$ \to \mathrm{e}^+\mathrm{e}^-$      &   1.5   &   $\approx 10^7$ &   $\approx 10^7$ LEP\\
W$ \to \mathrm{e}\nu$        &    15   &   $\approx 10^8$ &   $\approx 10^4$ LEP  \\
t$\bar{\mathrm{t}}$ &   800   &   $\approx 10^7$ &   $\approx 10^4$ Tevatron   \\
\hline
\end{tabular}
\end{center}
\end{table}

A detailed description of the ATLAS detector can be found in~\cite{tdr}.  
The inner-detector (ID), located inside the 2~T solenoid magnetic field, provides precision tracking of charged particles and secondary vertex finding in the pseudo-rapidity region $|\eta| < 2.5$. The liquid argon electromagnetic calorimeter provides coverage up to $|\eta| < 3.2$ with region of high granularity within $|\eta| < 2.5$. Combined with the hadronic calorimeter the limit of hermicity lies at $|\eta| = 4.9$. Finally, the outer muon spectrometer with an air core toroid system allows allows precise muon momentum measurements within $|\eta| < 2.7$. 

 The following sections present the prospects for measuring the W mass, probing triple gauge boson couplings and measuring the top quark mass at ATLAS during the first three years of LHC operation.

\section{Precision for the Measurement of the W Mass}
\label{subsec:wmass}

\subsection{Motivation}
\label{sec:wmass}

The present world average value of the W mass is $m_{\mathrm{W}}= 80.426 \pm0.034$~GeV~\cite{lepeew}. Within the SM $m_{\mathrm{W}}$ is related to other fundamental parameters through
\begin{equation}
m_{\mathrm{W}} = \sqrt{\left(\frac{\pi\alpha}{G_{\mathrm{F}}\sqrt{2}}\right)}\frac{1}{\sin\theta_{\mathrm{w}}\sqrt{1-\Delta R}},
\end{equation}
where $\alpha$ is the fine structure constant, $G_{\mathrm{F}}$ the Fermi constant and $\theta_{\mathrm{w}}$ the Weinberg angle.  The radiative corrections $\Delta R$ receive contributions from the square of the top mass, $m_{\mathrm{t}}^2$, and the logarithm of the Higgs mass, $\log m_{\mathrm{h}}$. Precision measurements of $m_{\mathrm{W}}$ together with $m_{\mathrm{t}}$ (refer to section~\ref{sec:top}) will therefore provide an important consistency check on the mass of the Higgs boson when it is discovered and help determine whether the observed Higgs is indeed that of the SM.

To achieve equal weights in a $\chi^2$ test, the precision on the W mass, $\Delta m_{\mathrm{W}}$, should be related to that on the top mass, $\Delta m_{\mathrm{t}}$, by $\Delta m_{\mathrm{W}} \approx 0.7 \times 10^{-2} \Delta m_{\mathrm{t}}$.
An anticipated uncertainty on the top quark mass of $\sim$2~GeV therefore demands a W mass precision of $\sim$15~MeV. Such measurements would constrain $m_{\mathrm{h}}$ to within 30\%.

\subsection{Measurement at ATLAS}

The principal method envisaged for measuring the W mass at the LHC employs the leptonic decay channels. The cross-section  for $\mathrm{pp} \to \mathrm{W} + \mathrm{X}$ with $\mathrm{W} \to \mathrm{l} \nu$ and $\mathrm{l} = \mathrm{e}, \nu$ is 30~nb, corresponding to $300\times 10^{6}$ events per low luminosity year. For selection, a single isolated charged lepton inside the region devoted to precision physics $(|\eta |< 2.4)$, significant missing transverse energy and minimal hadronic activity are required. 

The W mass is extracted from the distribution of the W transverse mass, $m_{\mathrm{T}}^{\mathrm{W}}$, given by
\begin{equation}
m_{\mathrm{T}}^{\mathrm{W}} = \sqrt{2p_{\mathrm{T}}^{\mathrm{l}} p_{\mathrm{T}}^{\nu}(1-\cos \Delta \phi)}
\end{equation}
in which $\phi$ is the azimuthal angle between the charged lepton and the system X which recoils against the W. The transverse momentum of the missing neutrino $p_{\mathrm{T}}^{\nu}$ has to be reconstructed from that of the lepton and the recoil.

\begin{figure}[!hpt]
\centerline{\epsfig{figure=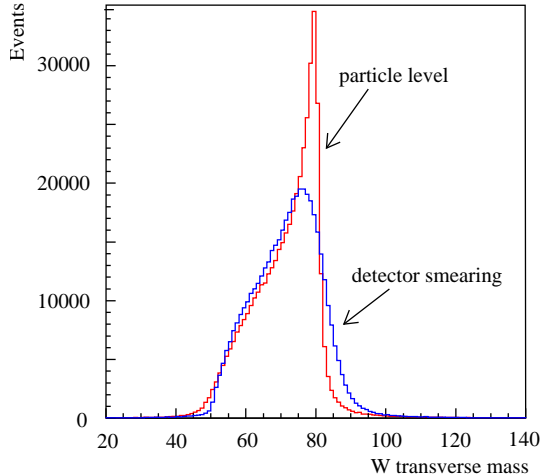,height=2.5in}}
\caption{W transverse mass distribution $m_{\mathrm{T}}^{\mathrm{W}}$ with and without detector smearing.}
\label{fig:wt}
\end{figure}

A simulated transverse mass distribution obtained using the ATLFAST~\cite{atlfast} fast detector simulation program for ATLAS is shown in figure~\ref{fig:wt}. 
The falling edge of the sharp Jacobian peak is sensitive to the W mass, though this sensitivity is reduced by the detector smearing. At high luminosity, this effect is significantly worsened by pile-up. The W mass can be extracted from the measured distribution by fitting to samples generated using different input values of $m_{\mathrm{W}}$.

\subsection{Expected Precision}

The large sample of W events available from the LHC means that the uncertainty on the W mass, $\Delta m_{\mathrm{W}}$, arising from the statistics is very small, around 2~MeV for 10~fb$^{-1}$ of data.  Table~\ref{tab:wmass} lists the various sources of systematic uncertainty together with their predicted contributions to $\Delta m_{\mathrm{W}}$.  Most of these have been evaluated by extrapolating from the experiences at the Tevatron based on the expected performance of the ATLAS detector~\cite{tdrw}.

Systematic uncertainties related to the physics knowledge enter through the ability to model the transverse mass distribution in the Monte Carlo. Factors include the modelling of the transverse momentum of the W, the transverse momentum of the recoil and radiative decays, the knowledge of the W width and parton density functions and the control of the backgrounds. 

The largest experimental uncertainty and the limiting factor governing the precision which will be reached on the W mass is the knowledge of the lepton energy and momentum ($E-p$) scale and resolution. The challenge is for ATLAS to achieve the necessary 0.02\% and 1.5\% precisions on the scale and resolution, respectively.  
{\it In-situ} calibrations, supplementing the initial hardware calibration and test-beam measurements, will play a crucial role in reaching this goal. These will make use, for example, of the large Z samples available, with Z$\to\mu\mu$ events being used for the calibration of the muon system and Z$\to$ee events for the electromagnetic calorimeter.
It is thought that for the ID and the electromagnetic calorimeter the required precision can be reached within one year of low luminosity running~\cite{epj}.
\begin{table}[t]
\caption{Sources of uncertainty in the measurement of the W mass at ATLAS and their contribution to $\Delta m_{\mathrm{W}}$ for each lepton species assuming 10~fb$^{-1}$ of data.}
\label{tab:wmass}
\vspace{0.2cm}
\begin{center}
\begin{tabular}{|l|c|c|}
\hline
Source             &       Assumptions         & $\Delta m_{\mathrm{W}}$  \\
\hline
Statistics         &        60M Ws reconstructed/year         &   $<$2~MeV               \\
\hline
W width            &                           &      7~MeV               \\
PDFs               &             &  $<$10~MeV               \\
Recoil modelling   &                           &      5~MeV               \\
Radiative Decays   &                           &  $<$10~MeV               \\
$p_{\mathrm{T}}^{\mathrm{W}}$ spectrum   &                           &      5~MeV               \\
Background understanding     &               & 5~MeV                    \\
\hline
Lepton identification &                        &      5~MeV               \\
Lepton $E-p$ scale   &    Known to 0.02\%        &  $<$15~MeV               \\
Lepton $E-p$ resolution & Known to 1.5\%         &      5~MeV               \\
\hline
Total              &                           &  $<$25~MeV               \\
\hline
\end{tabular}
\end{center}
\end{table}

Summing the contributions to $\Delta m_{\mathrm{W}}$ in quadrature, a total uncertainty of 25~MeV per channel is obtained: combining the electron and muon channels this is reduced to 20~MeV. This can be further reduced to the target of 15~MeV by combining with the results from CMS.

\section{Charged Triple Gauge Couplings}
\subsection{Motivation}

The local gauge symmetry of the SM generates self-interactions of the gauge fields, the form and strength of which are recorded in the term ${\mathbf W^{\mu\nu}} \cdot {\bf W_{\mu\nu}}$ of the Electroweak Lagrangian. This term leads to triple gauge couplings (TGCs) of the form WW$\gamma$ and WWZ and quartic gauge couplings (QGCs) WWWW, WWZZ, WW$\gamma\gamma$ and WWZ$\gamma$. Studying the form and structure of these gauge boson self-couplings is therefore an important test of the non-Abelian  $SU(2)_L \times U(1)_Y$ SM gauge structure.

  The SM TGCs have been beautifully confirmed at LEP but it remains possible that new physics at an unprobed energy scale may have low energy effects equivalent to anomalous TGCs (ATGCs), the study of which usually proceeds via an effective Lagrangian formalism. In the case of the charged triple gauge couplings, the most general Lorentz invariant effective Lagrangian for WWV where V $= \gamma, \mathrm{Z}$ contains 14 independent couplings, 7 for each vertex~\cite{tgclep}.
By requiring
electromagnetic gauge invariance (charge conservation) and $C$ and $P$
conservation the number of independent couplings is reduced to five, commonly written as $\{\Delta g_1^{\mathrm{Z}}, \Delta \kappa_{\mathrm{Z}}, \Delta \kappa_{\gamma}, \lambda_{\mathrm{Z}}, \lambda_{\gamma}\}$. In the SM these are all zero.  
The current combined LEP limits~\cite{tgclep} are given in table~\ref{tab:tgcs}: only three parameters appear as the LEP experiments impose additional symmetry constraints which further reduce the number of independent parameters.



\subsection{Measurement at ATLAS} 

The effective terms for the ATGC vertices are linear in the anomalous couplings, meaning that the cross-section for a process that receives a contribution from an anomalous vertex will vary quadratically with the coupling parameters. It is therefore possible to set limits on the ATGCs by simply comparing observed and expected event rates. However, it is desirable to plot the distribution of some observable which is sensitive to the different ATGCs.  This approach is less susceptible to uncertainties in the overall normalisation, for example from the luminosity, and moreover, should any deviation from the SM be indicated, helps to disentangle the different possible ATGC contributions.

\begin{figure}[!h]
\centerline{\epsfig{figure=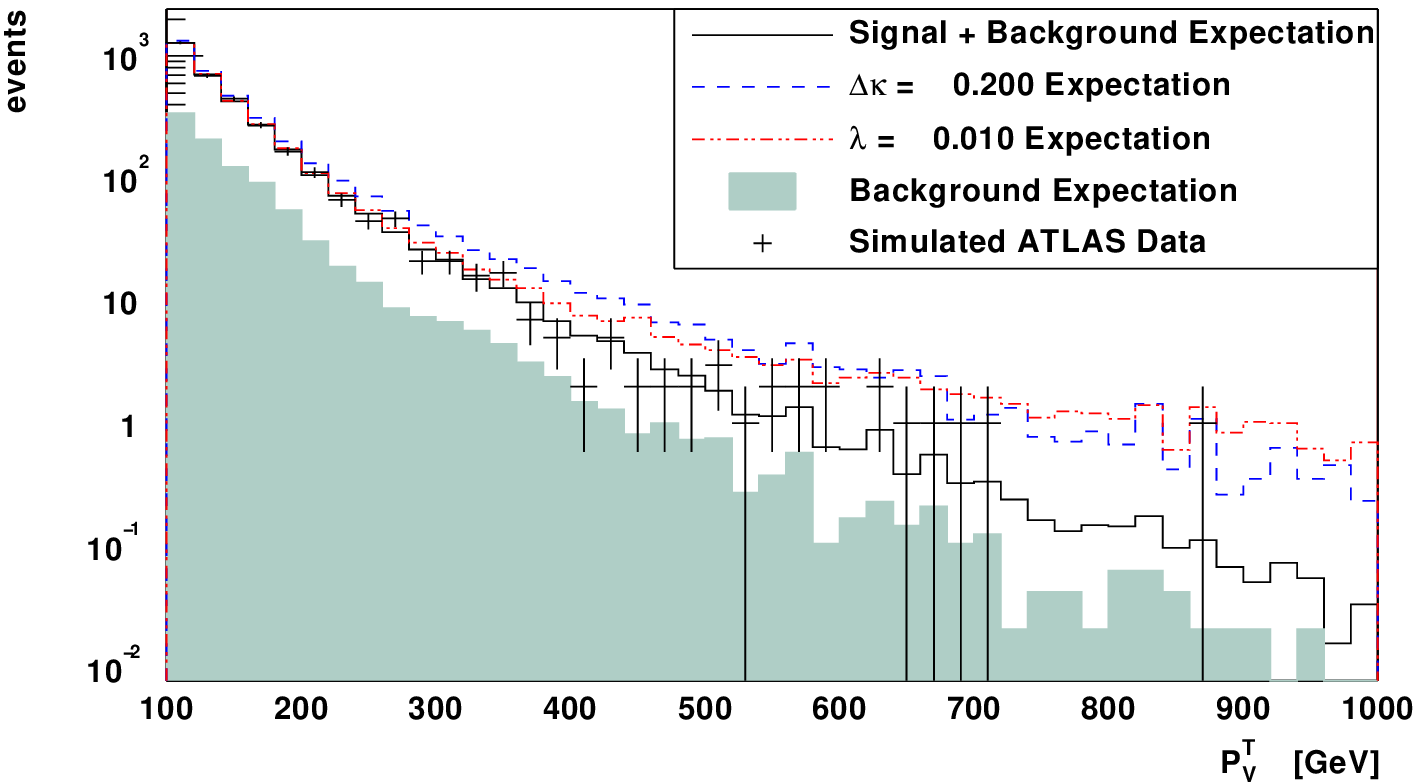,height=2.65in}}
\vskip 0.1in
\centerline{\epsfig{figure=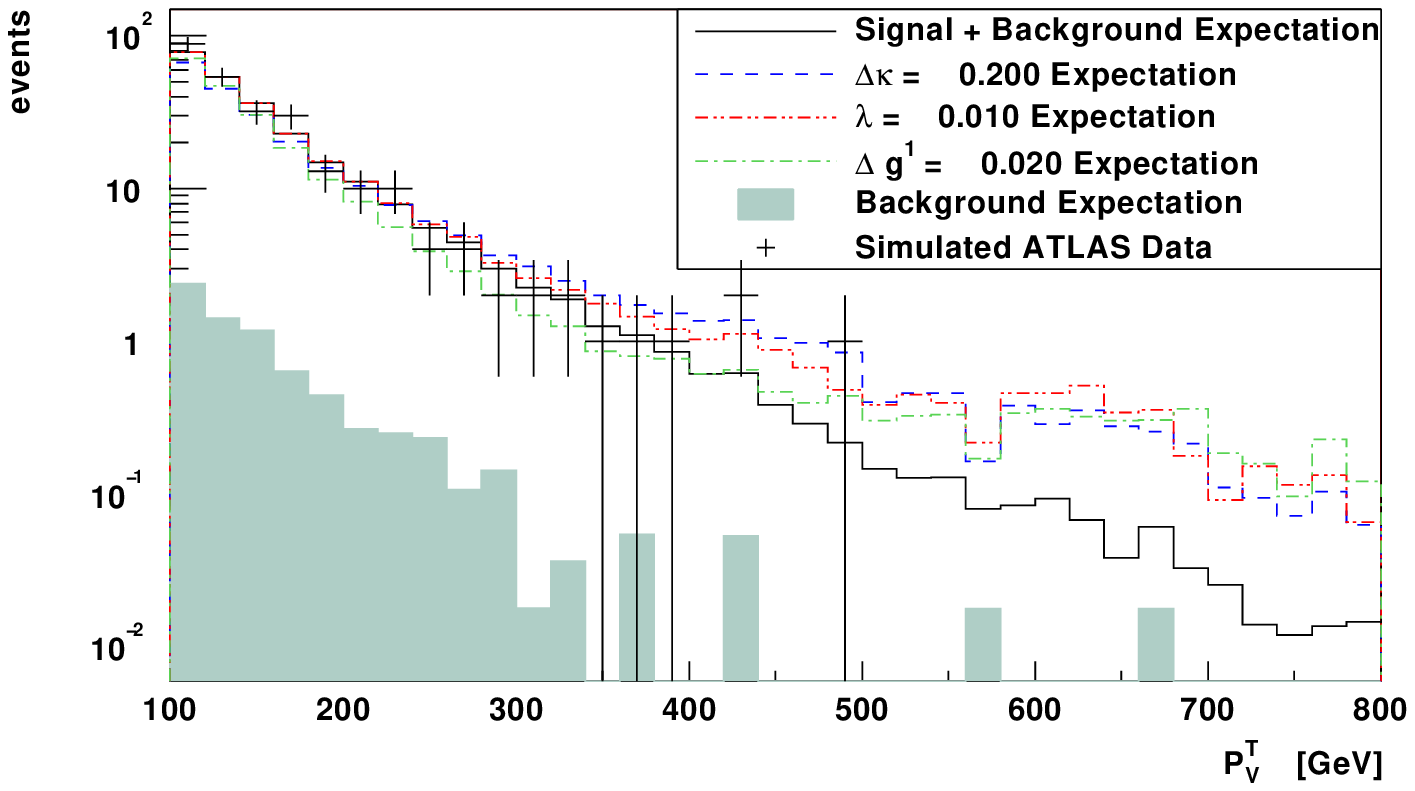,height=3.0in}}
\caption{Top: the transverse momentum of the photon in W$\gamma$ production. The points with error bars represent the data for a single ATLAS experiment of 30~fb$^{-1}$ using SM TGC parameters and including backgrounds. The lines show the reference distributions for the indicated choice of TGC parameters. The shaded area is the background and is independent of the TGCs. Bottom: as above but for the transverse momentum of the Z boson in WZ production.} 
\label{fig:wzgpt}
\end{figure}

Dobbs and Lefebvre have studied the possibilities of measuring the charged TGCs parameters $\kappa_{\gamma}$ and $\lambda_{\gamma}$ in the W$\gamma$ production process~\cite{matt1} and the parameters $g_1^{\mathrm{Z}}$, $\kappa_{\mathrm{Z}}$ and $\lambda_{\mathrm{Z}}$ in WZ production~\cite{matt2}.  In both cases the $p_{\mathrm{T}}$ spectrum of the photon or Z boson was found to offer the optimal sensitivity to the possible ATGCs. Examples of distributions obtained within the ATLFAST framework are shown in figure~\ref{fig:wzgpt}.  
For the W$\gamma$ process approximately 3000 events containing  a high $p_{\mathrm{T}}$ charged lepton, missing $p_{\mathrm{T}}$ and a high $p_{\mathrm{T}}$ photon satisfying $p_{\mathrm{T}}^{\gamma} > 100$~GeV are expected in 30~fb$^{-1}$ of data. For this selection the SM expectation for W$\gamma$ and non-W$\gamma$ backgrounds are small. For WZ production, around 1200 events are expected and again the backgrounds are small in the region $p_{\mathrm{T}}^{\mathrm{Z}}> 100$~GeV. 

Sensitivity to possible anomalous neutral TGCs of the form Z$\gamma\gamma$, ZZ$\gamma$ and ZZZ, which are absent in the SM, have also been studied in the Z$\gamma$~\cite{hass2} and ZZ production processes~\cite{hass1}.

\subsection{Expected Precision}

The expected limits which can be obtained on the charged ATGC parameters are presented in table~\ref{tab:tgcs}.  
They were found to be unitarity safe and are thus reported without any cutoff or form factor. The improvement over the current LEP limits is greatest for the $\lambda$-like couplings which enhance the cross-section by a factor proportional to the parton centre-of-mass energy squared, $\hat{s}$. For the $\kappa$-like couplings this enhancement goes only as $\sqrt{\hat{s}}$. 

These results are both systematics and statistics limited due to the cut on the photon or Z transverse momentum required to boost sensitivity to the ATGCs. The systematic effects arise from the modelling of the background, the knowledge of the parton density functions and knowledge of the detector smearing within ATLFAST.

\begin{table}[t]
\caption{Limits at 95\% confidence level on the charged ATGC parameters from the combined LEP results and expected in 30~fb$^{-1}$ of ATLAS data.}
\label{tab:tgcs}
\vspace{0.4cm}
\begin{center}
\begin{tabular}{|c|c|c|}
\hline
ATGC Parameter            &          95\% CL LEP2          &  95\% CL 30~fb$^{-1}$ ATLAS \\
\hline
$\Delta \kappa_{\gamma}$   &     -0.105, +0.069        & -0.075, +0.076       \\
$\lambda_{\gamma}$         &     -0.059, +0.026        & -0.0035, +0.0035     \\
\hline
$\Delta g_1^{\mathrm{Z}}$             &     -0.051,+0.034        & -0.0086, +0.011      \\
$\Delta \kappa_{\mathrm{Z}}$          &                          & -0.11, +0.12         \\
$\lambda_{\mathrm{Z}} $               &                          & -0.0072, +0.0072     \\
\hline
\end{tabular}
\end{center}
\end{table}


\section{Top Mass Precision}
\label{sec:top}

\subsection{Motivation}

The current world average value of mass of the top quark is $m_{\mathrm{t}} = 178.0 \pm 4.3$~GeV based on results from the Tevatron~\cite{topmass}.  The top quark mass is a fundamental parameter of the SM; moreover, combined with measurements of $m_{\mathrm{W}}$ it provides a consistency check on the SM Higg's mass as described in section~\ref{sec:wmass}.

\subsection{Measurement at ATLAS}

\begin{figure}[hpt!]
\centerline{\epsfig{figure=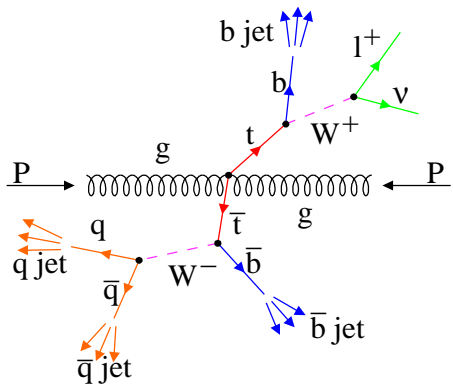,height=2.0in}}
\caption{The topology of $\mathrm{t\bar{t}}$ production at the LHC through the lepton plus jets channel.}
\label{fig:top}
\end{figure}

At low luminosity the next-to-leading order cross-section calculation predicts a rate of $8\times 10^6$ t$\bar{\mathrm{t}}$ events per low luminosity year at the LHC, 90\% of which are produced through $\mathrm{gg}\to \mathrm{t\bar{t}}$~\cite{topq}. 
Since in the SM the t decays almost exclusively to Wb, the $\mathrm{t\bar{t}}$ events are categorised by the subsequent decay of the two Ws into three types: multijet events where both W decays are hadronic, lepton plus jets events and di-lepton events.  The lepton plus jets channel (figure~\ref{fig:top}) 
represents the most favourable route to measuring the t quark mass, though alternative methods have been discussed (see for example~\cite{topq}).  Considering electrons and muons the branching ratio for the dilepton channel is just 5\%, compared to 30\% for the lepton plus jets channel which, containing just one missing neutrino, can also be fully reconstructed within a quadratic ambiguity. 
The multijet channel has a higher branching ratio but suffers from a large QCD background and the lack of the trigger efficiency which accompanies an isolated charged lepton.

The single lepton plus jets selection requires an isolated lepton plus missing energy and four high energy jets, at least two of which are b-tagged. The efficiency (3.5\%) can be improved by relaxing the b-tagging requirement to just one jet at the cost of decreasing the purity of the selected sample. Given that the analysis with two b-tagged jets will not be statistics limited the tighter selection can be maintained.

\begin{figure}[!bt]
\centerline{\epsfig{figure=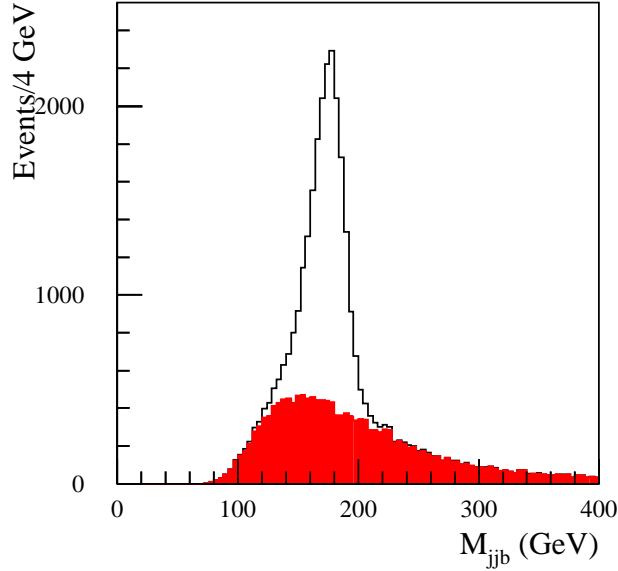,height=3.0in}}
\caption{Top quark mass distribution obtained using ATLFAST for 10~fb$^{-1}$ of data. The shaded background is combinatorial.}
\label{fig:topplot}
\end{figure}

The mass measurement proceeds by reconstructing the Wjj system from the two jets which are not b-tagged and combining with one of the b-tagged jets to reconstruct the jjb system. The jjb invariant mass distribution is plotted in figure~\ref{fig:topplot}~\cite{topq} together with the combinatorial background. The top quark mass is extracted from this plot by fitting to the peak using Monte Carlo samples generated with different values of $m_{\mathrm{t}}$.

\subsection{Expected Precision}

The large sample of t$\bar{\mathrm{t}}$ events available from the LHC means that the uncertainty on the t quark mass, $\Delta m_{\mathrm{t}}$, arising from the statistics is very small, around 0.1~GeV for 10~fb$^{-1}$ of data.  Table~\ref{tab:tmass} lists the various sources of systematic uncertainty together with their predicted contributions to $\Delta m_{\mathrm{t}}$.  
The main experimental uncertainty is the calibration of jet energy scale: it has been assumed that this will be known to 1\%. {\it In-situ} calibrations will be performed using a clean sample of W$\to$jj events.

The largest physics uncertainty arises from the modelling of the final state radiation. The effect of this uncertainty can be reduced by employing a slightly modified method in which a kinematic fit is applied, constraining the jjb invariant mass to that of the system containing the other b quark, the charged lepton and the neutrino, $m_{\mathrm{l}\nu\mathrm{b}}$.

Summing the contributions to $\Delta m_{\mathrm{t}}$ in quadrature, ATLAS should attain a precision of 1.3~GeV on the top quark mass 
in one year of low luminosity running.
\begin{table}[t]
\caption{Sources of uncertainty and their contribution to the t quark mass precision for 10~fb$^{-1}$ of ATLAS data.}
\label{tab:tmass}
\vspace{0.4cm}
\begin{center}
\begin{tabular}{|l|c|}
\hline
Source               &          $\Delta m_{\mathrm{t}}$ \\
\hline
Statistics           &           0.1~GeV     \\
\hline
b fragmentation      &           0.1~GeV     \\
Initial state radiation                  &           0.1~GeV     \\
Final state radiation                 &           1.0~GeV     \\
Background           &           0.1~GeV     \\
\hline
Light quark jet energy scale calibration   & 0.2~GeV \\
b quark jet energy scale calibration       & 0.7~GeV \\
\hline
Total                &           $\approx$ 1.3~GeV  \\
\hline
\end{tabular}
\end{center}
\end{table}


\section{Summary}

The LHC is a factory for many Electroweak processes and combined with its high centre-of-mass energy will enable many precise tests of the SM.
The high statistics means that the limit on the precision measurements will be often be systematics dominated, but in turn the large physics samples which are available can be used to constrain these uncertainties.  It remains vital however that the behaviour of ATLAS detector is well understood if the potential of the LHC is to be maximised.

\section*{Acknowledgments}

This report reflects the work of many members of the ATLAS collaboration who have also made use of the physics analysis framework and tools which are the result of collaboration-wide efforts.  I should like to thank Matt Dobbs and Craig Buttar for their help in preparing my Moriond presentation and not least the organisers of the XXXIXth Rencontres de Moriond for an informative an enjoyable conference.

\end{document}